\documentclass[12pt,thmsa]{article}
\usepackage{amsfonts}
\usepackage{graphicx}
\usepackage{epsfig}
\usepackage{graphics}

\makeatletter
\newcommand{\row}[1]{\mathord{\buildrel{\lower3pt\hbox{$\scriptscriptstyle\rightarrow$}}\over #1}}
\newcommand{\col}[1]{{#1^{\raisebox{2pt}[\height]{$\scriptstyle\downarrow$}}}}
\newcommand{\dyadic}[1]{\mathord{\dyadic@rrow{#1}}}
\newcommand{\dyadic@rrow}[1]{
\begin{picture}(12,12)(-1,0)
\put(-2,12){\makebox(0,0)[t]{$\scriptscriptstyle\downarrow$}}
\put(-2,12){\makebox(0,0)[l]{$\scriptscriptstyle\longrightarrow$}}
\put(5,0){\makebox(0,0)[b]{$#1$}}
\end{picture}
}

\newcommand{\ket}[1]{\bigl| #1 \bigr\rangle}

\input{tcilatex}
\begin{document}
\begin{center}
\textbf{\Large Dynamics of information in the presence of
deformation}

N. Metwally \\[0pt]

Math. Dept., Faculty of Science, South Valley University, Aswan, Egypt. \\[0pt]
E.mail: Nmetwally$@$gmail.com
\end{center}

\begin{abstract}
The entanglement of atomic system consists of two atoms interacts
with a deformed cavity mode is quantified by the means of Bloch
vectors and the cross dyadic of the traveling state inside the
cavity. For large value of the deformation the amplitude of Bloch
vectors decrease very fast and consequently the traveling state
turns into mixed state quickly. The generated entangled state is
used as quantum channel to implement quantum teleportation protocol.
It is shown that both of the deformed parameter and the number of
photons inside the cavity play a central role on controlling the
fidelity of the transmitted information.

\end{abstract}

\section{Introduction}

Controlling quantum coherence is one of the most fundamental
issues in  quantum information processing \cite{nil}.  From a
practical  point of view the  investigating  of the behavior of
entanglement in the presences of noise is very important, where
there are many studies that has been done in different directions.
Among these noise is the deformation, which arise from the defects
of devices \cite{Lav,Abou,Bon}. So, in the presence of
deformation, one can generate a deformed entangled state  and
consequently there will be a deformation on the information
processing.

The quantization of the field in the presence of deformed
operators has been investigated  in different topics, where the
relation between the deformed and non deformed operators have been
obtained in \cite{Hong,Man,Lav2}. As an applications of the
deformed operators, Mancini and  Man'ko have  compared the
dynamics of information  which is coded in deformed cat state with
non-deformed cat state \cite{Man2}. Also, in \cite{Yuk} the
holonomic quantum gates using isospectral deformations of an Ising
model  have been constructed. The survival of  quantum
coherence  against dissipation provided to superpose
distinguishable coherent states of suitable deformed field has
been investigated in \cite{Man3}. Recently, Metwally and et.al,
have showed that the non-classical properties of a two qubits
state can be enhanced via deformed operators\cite{Metwally}.

These efforts motivate us to introduce a different application of
the deformation. In this work, we consider a system consists of of
two separable  atoms  interacts with a cavity mode with
multiphotons. The operators which describe the field are assumed
to be deformed. We investigate the behavior of  entanglement which
is generated between the two atoms. Also, the dynamics of the
Bloch vectors are investigated for this deformed atomic system.
Finally, we use the generated entangled state between the two
atoms as quantum channel to implement the quantum teleportation
protocol.
 The effect of the deformation on the fidelity of the transmitted
 information is investigated.

 This paper is organized as follows: In Sec.2, we introduce the
 model and its solution by means of the Bloch vectors and the
 cross dyadic.  The entanglement and the dynamics of the Bloch
 vectors are studied in Sec.3. The quantum
 teleportation protocol is the subject of Sec.4. Finally, we summarize our results in Sec.5,

\section{ Model}
Let us assume that we have   atomic system consists of two atoms
 interacts with cavity mode with $m$ \ photons. The annihilation
and creation operators which describe the cavity mode are assumed
to be deformed. The deformed operators result as a distortion of
the usual annihiliation and creation operators \cite{Lav,Man}.  In
the rotating wave approximation, the interaction of
 the cavity mode and the atomic systems is described by the
Hamilation ,
\begin{equation}
\hat{H}\mathbf{=\varpi }_{0}{a_q}^{\dag
}a_q+\frac{1}{2}\omega(\sigma_z+\tau_z)+\lambda_1 (\sigma_{+}a_q
^{m}+\sigma a_q ^{\dag m})+ \lambda_2 (\tau_{+}a_q ^{m}+\tau_{-}
a_q ^{\dag m}) , \label{Ham}
\end{equation}%
where, $\varpi _{0}$ is the frequencies of the field, $\omega_i$ are
the atomic transition frequency, $\lambda_i,i=1,2 $ are the
coupling constants between the atoms and the field, $ a_q$ and \
$a_q^{\dag }$ are  deformed annihilation and creation operators,
which can be described by means of the non-deformed operators $a $
and $a^{\dagger}$ as,
\begin{equation}
a_q=af(n),\quad a_q^{\dag }=a^{\dag }f(n),  \label{DefOper}
\end{equation}%
where $f(\hat{n})$\ is a function of the number of photons $\bar
n=a^{\dag }a.$ The deformed operators  $a_q$ and \ ${a_q}^{\dag }$
satisfy the commutation relations,
\begin{eqnarray}
\left[ a_q,{a_q}^{\dag }\right] &=&(\hat{n}+1)f^{2}(\hat{n}+1)-\hat{n}f^{2}(\hat{n}),  \label{DefRel} \\
\left[ a_q,n\right] &=&a_q,~\left[ a_q^{\dag },n\right] =-a_q
^{\dag }.  \nonumber
\end{eqnarray}%
The operators $\sigma_{\pm }=\sigma_x\pm i\sigma_y$, $\tau_{\pm
}=\tau_x\pm i\tau_y$
 and
$\sigma_{z}$,$\tau_z$ are the raising (lowering) and inversion
operators for the two atoms. The operators $\sigma_i$ and $\tau_i,
i=x,y,z$ are the Pauli's operators for the first and the second
atoms respectively. In this treatment, the function $f(\hat{n})$
represents what is called the $q-$deformation  and it is defined as
\begin{equation}
f(\hat{n})=\sqrt{\frac{1-q^n}{n(1-q)}}.
\end{equation}

In the invariant sub-space of the global system, we can assume a
set of
complete basis of the field-atomic system as $\left\vert e,e,n\right\rangle $%
, $\ \left\vert e,g,n+2\right\rangle ,$ $\left\vert
g,e,n+2\right\rangle $ \ and $\left\vert g,g,n+2\right\rangle.$
Now, let  the \ initial state of the atomic system, $\left\vert
\psi _{a}(0)\right\rangle =a_{1}\left\vert ee\right\rangle
+a_{2}\left\vert eg\right\rangle +a_{3}\left\vert ge\right\rangle
+a_{4}\left\vert gg\right\rangle ,$ while the \ field is initially
prepared in the coherent state $\left\vert \psi
_{f}(0)\right\rangle =\sum_{n=0}^{\infty }W_{n}\left\vert
n\right\rangle
,W_{n}=\frac{\overline{n}^{n}}{\sqrt{n!}}Exp\left( -\frac{1}{2}|\overline{n}%
|^{2}\right).$

For a convenient notations  the density operator of the atomic system is
described by the  Bloch vectors and the cross dyadic as
\cite{metwally1,Englert2},
\begin{equation}
\rho _{a}(0)=\frac{1}{4}\left( +\row{s}_{0}\cdot\col\sigma  +
\row{t}_{0}\cdot\col\tau+\row\sigma\cdot\dyadic{C}(0)\cdot\col\tau\right)
,  \label{Inatom}
\end{equation}%
where,
 \begin{equation}
\row{s}_{0}=(s_{x}(0),s_{y}(0),s_{z}(0)),\quad
\row{t}_{0}=(t_{x}(0),t_{y}(0),t_{z}(0)),
\end{equation}
 are the initial Bloch vectors of the two atoms respectively, with
\begin{eqnarray}
s_{x}(0) &=&(a_{1}a_{3}^{\ast }+a_{3}a_{1}^{\ast })+(a_{2}a_{4}^{\ast
}+a_{4}a_{2}^{\ast }),s_{y}(0)=i(a_{3}a_{1}^{\ast }-a_{1}a_{3}^{\ast
})+i(a_{4}a_{2}^{\ast }-ia_{2}a_{4}^{\ast }),  \nonumber \\
s_{z}(0)
&=&|a_{1}|^{2}+|a_{2}|^{2}-|a_{3}|^{2}-|a_{4}|^{2},t_{x}(0)=(a_{1}a_{2}^{%
\ast }+a_{2}a_{1}^{\ast })+(a_{3}a_{4}^{\ast }+a_{4}a_{3}^{\ast }),
\nonumber \\
t_{y}(0) &=&i(a_{2}a_{1}^{\ast }-a_{1}a_{2}^{\ast })+i(a_{4}a_{3}^{\ast
}-a_{3}a_{4}^{\ast
}),t_{z}(0)=|a_{1}|^{2}-a_{2}|^{2}+|a_{3}|^{2}-|a_{4}|^{2}.  \label{BlochIni}
\end{eqnarray}%
and the elements of the   initial cross dyadic between the two
atoms are given by,
\begin{eqnarray}
c_{xx}(0) &=&a_{1}a_{4}^{\ast }+a_{2}a_{3}^{\ast }+a_{3}a_{2}^{\ast
}+a_{4}a_{1}^{\ast },c_{xy}(0)=i(a_{4}a_{1}^{\ast }-a_{1}a_{4}^{\ast
})+i(a_{2}a_{3}^{\ast }-a_{3}a_{2}^{\ast }),  \nonumber \\
c_{xz}(0) &=&(a_{1}a_{3}^{\ast }+a_{3}a_{1}^{\ast })-(a_{2}a_{4}^{\ast
}+a_{4}a_{2}^{\ast }),c_{yx}(0)=i(a_{4}a_{1}^{\ast }-a_{1}a_{4}^{\ast
})+i(a_{3}a_{2}^{\ast }-a_{2}a_{3}^{\ast }),  \nonumber \\
c_{yy}(0) &=&(a_{3}a_{2}^{\ast }+a_{2}a_{3}^{\ast })-(a_{4}a_{1}^{\ast
}+a_{1}a_{4}^{\ast }),c_{yz}(0)=i(a_{2}a_{4}^{\ast }-a_{4}a_{2}^{\ast
})+i(a_{3}a_{1}^{\ast }-a_{1}a_{3}^{\ast }),  \nonumber \\
c_{zx}(0) &=&(a_{1}a_{2}^{\ast }+a_{2}a_{1}^{\ast })-(a_{3}a_{4}^{\ast
}+a_{4}a_{3}^{\ast }),c_{zy}(0)=i(a_{2}a_{1}^{\ast }-a_{1}a_{2}^{\ast
})+i(a_{3}a_{4}^{\ast }-a_{4}a_{3}^{\ast }),  \nonumber \\
c_{zz}(0) &=&|a_{1}|^{2}-|a_{2}|^{2}-|a_{3}|^{2}+|a_{4}|^{2}.
\label{IniDyad}
\end{eqnarray}
Also, the  density operator of  the field is given by,
\begin{equation}
\rho _{f}(0)=\sum_{n=0}^{\infty }W_{n}^{2}\left\vert
n\right\rangle \left\langle n\right\vert.  \label{InField}
\end{equation}
Then using the Hamiltonian (\ref{Ham}), the initial state of the
atomic system (\ref{Inatom}), and the initial state of the field
(\ref{InField}) one gets the time evolution of the density
operator of the field and the atomic system, where
$\lambda_1=\lambda_2=\lambda$, $\omega_1=\omega_2=\omega$, and
$\omega_0-\omega=\Delta=0$. After tracing out the state of the
field, the density operator  of the atomic system is given by
\begin{equation}
\rho _{a}(t)=\frac{1}{4}\left( 1+\mathord{\buildrel{\lower3pt\hbox{$%
\scriptscriptstyle\rightarrow$}}\over s}\cdot{\sigma ^{\raisebox{2pt}[%
\height]{$\scriptstyle\downarrow$}}}+ \mathord{\buildrel{\lower3pt\hbox{$%
\scriptscriptstyle\rightarrow$}}\over t}\cdot{\tau^{\raisebox{2pt}[%
\height]{$\scriptstyle\downarrow$}}}+\mathord{\buildrel{\lower3pt\hbox{$%
\scriptscriptstyle\rightarrow$}}\over \sigma }\cdot\mathord{\dyadic@rrow{C}}%
\cdot{\tau^{\raisebox{2pt}[\height]{$\scriptstyle\downarrow$}}} \right) ,
\label{FinalAtomic}
\end{equation}
where,
\begin{eqnarray}
s_{x}(t) &=&\sum_{n=0}^{\infty }\left( c_{n}^{(1)}c_{n-2}^{\ast
(3)}+c_{n}^{(2)}c_{n-2}^{\ast (4)}+c_{n}^{(3)}c_{n+2}^{\ast
(1)}+c_{n}^{(4)}c_{n+2}^{\ast (2)}\right) ,  \nonumber \\
s_{y}(t) &=&\sum_{n=0}^{\infty }\left( -ic_{n}^{(1)}c_{n-2}^{\ast
(3)}-ic_{n}^{(2)}c_{n-2}^{\ast (4)}+ic_{n}^{(3)}c_{n+2}^{\ast
(1)}+ic_{n}^{(4)}c_{n+2}^{\ast (2)}\right) ,  \nonumber \\
s_{z}(t) &=&\sum_{n=0}^{\infty }\left(
|c_{n}^{(1)}|^{2}+|c_{n}^{(2)}|^{2}-|c_{n}^{(3)}|^{2}-|c_{n}^{(4)}|^{2}%
\right) ,  \nonumber \\
t_{x}(t) &=&\sum_{n=0}^{\infty }\left( c_{n}^{(1)}c_{n-2}^{\ast
(2)}+c_{n}^{(2)}c_{n+2}^{\ast (1)}+c_{n}^{(3)}c_{n-2}^{\ast
(4)}+c_{n}^{(4)}c_{n+2}^{\ast (3)}\right) ,  \nonumber \\
t_{y}(t) &=&\sum_{n=0}^{\infty }\left( -ic_{n}^{(1)}c_{n-2}^{\ast
(2)}+ic_{n}^{(2)}c_{n+2}^{\ast (1)}-ic_{n}^{(3)}c_{n-2}^{\ast
(4)}+ic_{n}^{(4)}c_{n+2}^{\ast (3)}\right) ,  \nonumber \\
t_{z}(t) &=&\sum_{n=0}^{\infty }\left(
|c_{n}^{(1)}|^{2}-|c_{n}^{(2)}|^{2}+|c_{n}^{(3)}|^{2}-|c_{n}^{(4)}|^{2}
\right) ,  \label{FinalBloch}
\end{eqnarray}
and the elements of  the cross dyadic are,
\begin{eqnarray}
c_{xx}(t) &=&\sum_{n=0}^{\infty }\left( c_{n}^{(1)}c_{n-4}^{\ast
(4)}+c_{n}^{(4)}c_{n+4}^{\ast (1)}+c_{n}^{(2)}c_{n}^{\ast
(3)}+c_{n}^{(3)}c_{n}^{\ast (2)}\right) ,  \nonumber \\
c_{xy}(t) &=&i\sum_{n=0}^{\infty }\left( -c_{n}^{(1)}c_{n-4}^{\ast
(4)}+c_{n}^{(4)}c_{n+4}^{\ast (1)}+c_{n}^{(2)}c_{n}^{\ast
(3)}-c_{n}^{(3)}c_{n}^{\ast (2)}\right) ,  \nonumber \\
c_{xz}(t) &=&\sum_{n=0}^{\infty }\left( c_{n}^{(1)}c_{n-2}^{\ast
(3)}+c_{n}^{(3)}c_{n+2}^{\ast (1)}-c_{n}^{(2)}c_{n-2}^{\ast
(4)}+c_{n}^{(4)}c_{n+2}^{\ast (2)}\right) ,  \nonumber \\
c_{yx}(t) &=&i\sum_{n=0}^{\infty }\left( -c_{n}^{(1)}c_{n-4}^{\ast
(4)}+c_{n}^{(4)}c_{n+4}^{\ast (1)}-c_{n}^{(2)}c_{n}^{\ast
(3)}+c_{n}^{(3)}c_{n}^{\ast (2)}\right) .  \nonumber \\
c_{yy}(t) &=&-\sum_{n=0}^{\infty }\left( c_{n}^{(1)}c_{n-4}^{\ast
(4)}+c_{n}^{(4)}c_{n+4}^{\ast (1)}-c_{n}^{(2)}c_{n}^{\ast
(3)}-c_{n}^{(3)}c_{n}^{\ast (2)}\right) ,  \nonumber \\
c_{yz}(t) &=&i\sum_{n=0}^{\infty }\left( -c_{n}^{(1)}c_{n-2}^{\ast
(3)}+c_{n}^{(3)}c_{n+2}^{\ast (1)}+c_{n}^{(2)}c_{n-2}^{\ast
(4)}+c_{n}^{(4)}c_{n+2}^{\ast (2)}\right) ,  \nonumber \\
c_{zx}(t) &=&\sum_{n=0}^{\infty }\left( c_{n}^{(1)}c_{n-2}^{\ast
(2)}+c_{n}^{(2)}c_{n+2}^{\ast (1)}-c_{n}^{(3)}c_{n-2}^{\ast
(4)}-c_{n}^{(4)}c_{n+2}^{\ast (3)}\right) ,  \nonumber \\
c_{zy}(t) &=&i\sum_{n=0}^{\infty }\left( -c_{n}^{(1)}c_{n-2}^{\ast
(2)}+c_{n}^{(2)}c_{n+2}^{\ast (1)}+c_{n}^{(3)}c_{n-2}^{\ast
(4)}-c_{n}^{(4)}c_{n+2}^{\ast (3)}\right) ,  \nonumber \\
c_{zz}(t) &=&\sum_{n=0}^{\infty }\left(
|c_{n}^{(1)}|^{2}-|c_{n}^{(2)}|^{2}-|c_{n}^{(3)}|^{2}+|c_{n}^{(4)}|^{2}%
\right),  \label{Finaldyd}
\end{eqnarray}
with,
\begin{eqnarray}
c_{n}^{(1)}(t) &=&a_{1}W_{n}-\nu _{1}(a_{1}\nu _{1}W_{n}+a_{4}\nu
_{2}W_{n+2})\frac{\sin ^{2}\mu _{n} t}{\mu _{n}^{2}}-i\nu
_{1}(a_{2}+a_{3})W_{n+1}\frac{\sin 2\mu _{n} t}{2\mu _{n} t},
  \nonumber \\
c_{n}^{(2)}(t) &=&W_{n+1}(a_{2}\cos ^{2}\mu _{n} t-a_{3}\sin
^{2}\mu _{n} t)-i(a_{1}\nu _{1}W_{n}+a_{4}\nu
_{2}W_{n+2})\frac{\sin 2\mu _{n} t}{2\mu _{n}t},
\nonumber \\
c_{n}^{(3)}(t) &=&W_{n+1}(a_{3}\cos ^{2}\mu _{n} t-a_{2}\sin
^{2}\mu _{n} t)-i(a_{1}\nu _{1}W_{n}+a_{4}\nu
_{2}W_{n+2})\frac{\sin 2\mu _{n}t}{2\mu _{n}t},
\nonumber \\
c_{n}^{(4)}(t) &=&a_{4}W_{n+2}-\nu _{2}(a_{1}\nu
_{1}W_{n}+a_{4}\nu _{2}W_{n+2})\frac{\sin ^{2}\mu _{n} t}{\mu
_{n}^{2}}-i\nu _{2}(a_{2}+a_{3})W_{n+1}\frac{\sin 2\mu _{n}t}{2\mu
_{n} t},\nonumber\\
\end{eqnarray}
where, $\mu_n=\frac{1}{\sqrt{2}}\sqrt{\nu_1^2(n)+\nu_2^2(n)}$,
$\nu_1(n)=\lambda\sqrt{(n+m)!/n!}$ and
$\nu_2(n)=\lambda\sqrt{(n+2m)!/(n+m)!}$.
\section{Dynamics of Entanglement}
Since we use a measure of entanglement depends on the Bloch
vectors  and the   cross dyadic of the two atoms, it is important
to shed light on the dynamics of these vectors.
 This study gives us a perception in the form of the state of the two  atoms  as it
 passes through the deformed cavity. We assume that the two atoms are identical, so the dynamics of
$|\row{s}|$ for the first atom and $|\row{t}|$ for the second atom
has the same behavior.

 Fig.(1) shows the dynamics of the Bloch vectors
 for different values of the deformation.
 For free deformation cavity, the amplitude of the Bloch vectors
 decreases as $\lambda t$ increases. Due to the
 flocculating behavior, the travelling state through the cavity turns
 into mixed and pure state several time.  As one consider the
 deformation,  the  amplitude of the Bloch vectors decrease
 more and the minimum values are always smaller than that depicted for the  free
 deformation case (dot-curves). However  as one increases the deformed parameter ($%
q=0.9)$, the amplitude of the Bloch vectors decrease faster as
shown in Fig.$(1a)$. This means that the state of the  two qubits
turns into a mixed state faster for larger values of the
deformity.

In Fig.(1b), the number of photons inside the cavity  is increased
($m=2$). It is clear that  for free deformation the amplitude of
the Bloch vectors oscillates so fast and increases as time
increases. So, by increasing the number of photons inside the
cavity, one can increase the purity of the travelling state. In
the presence of deformation ($q=0.5)$, the behavior of the
amplitude of the Bloch vectors is similar to that depicted for the
free deformation case. However as one increases the deformity
parameter more $(q=0.9)$, the amplitude of the Bloch vectors
increases. From Fig.(1a) and Fig(1b), it is clearly  that by
increasing the number of photons inside the cavity, one can avoid
the defected which results from the deformed cavity.

On the other hand, when the Bloch vectors vanish and the
entanglement still survival between the two atoms,
 this means that  an entangled   state of Werner type  \cite{Werner,Englert2} has been
  generated and  it can be written as:
  \begin{equation}
  \rho_w=\frac{1}{4}(1+x_1\sigma_x\tau_x+x_2\sigma_y\tau_y+x_3\sigma_z\tau_z).
  \end{equation}
  However, at certain  time say (t=2.5,5) as shown in Fig.(1a), the amplitudes of
  the Bloch vectors $|\row{s}|=|\row{t}|=1$ and the entanglement
  vanishes (see Fig.(2a)), this means that the atomic system turns into itsd initial
  state.

\begin{figure}
\begin{center}
\includegraphics[width=18pc,height=12pc]{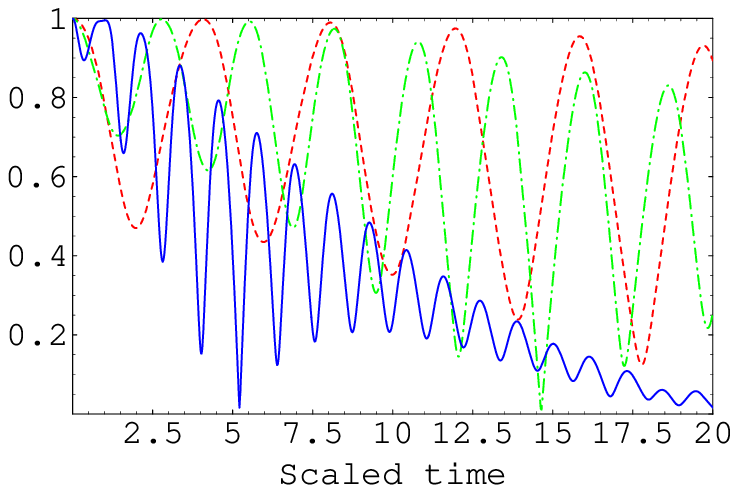} %
\includegraphics[width=18pc,height=12pc]{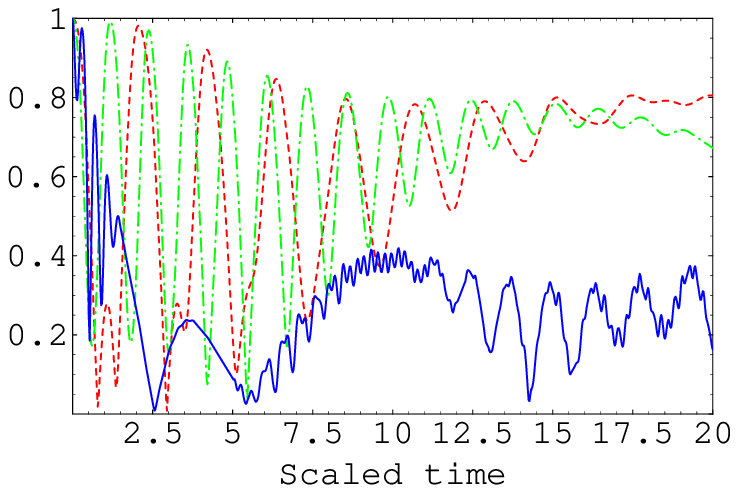}
\caption{The Bloch vector for the travailing state
(\ref{InitAstate}), where
 the deformed parameter $q=0, 0.5, 0.9$ and the mean photon number,
  $\bar n=10$ for the dot, dash -dot and solid curves,respectively (a)
$m=1$ and (b)$m=2$.  }
\end{center}
\end{figure}

To quantify the degree of entanglement between the two atoms, we
 use a measure  defined by means of the Bloch vectors and the cross
 dyadic. The entangled dyadic  is defined as
             \begin{equation}
             \dyadic{E}=\dyadic{C}-\row{s}\col{t}
                   \end{equation}
             where $\dyadic{C}$ is a $3\times 3$ matrix which is defined
             by(\ref{Finaldyd}) and $\row{s}\col{t}$ is also  a $3\times 3$ matrix whose elements can be obtained from
             (\ref{FinalBloch}). The degree of entanglement is
             defined as
            in \cite{Englert, Metwally2}
    \begin{equation}\label{DoE}
             \mathcal{E}=tr\{~{\mathord{\dyadic@rrow{E}}}^{\mathsf{T}}\cdot \mathord{%
             \dyadic@rrow{E}}\},
             \end{equation}
 where ${\mathord{\dyadic@rrow{E}}}^{\mathsf{T}}$ is the transpose matrix of
  the dyadic $\mathord{\dyadic@rrow{E}}$ and $\mathcal{E}=0$ for separable
             states.

Fig $(2$), describes the dynamics of entanglement which is
generated between  two atoms prepared initially in  a separable
state define by,
\begin{equation}\label{InitAstate}
\rho _{a}=\frac{1}{4}(1+s_{z}+t_{z}+c_{zz}).
\end{equation}%
\begin{figure}[t!]
\begin{center}
\includegraphics[width=18pc,height=12pc]{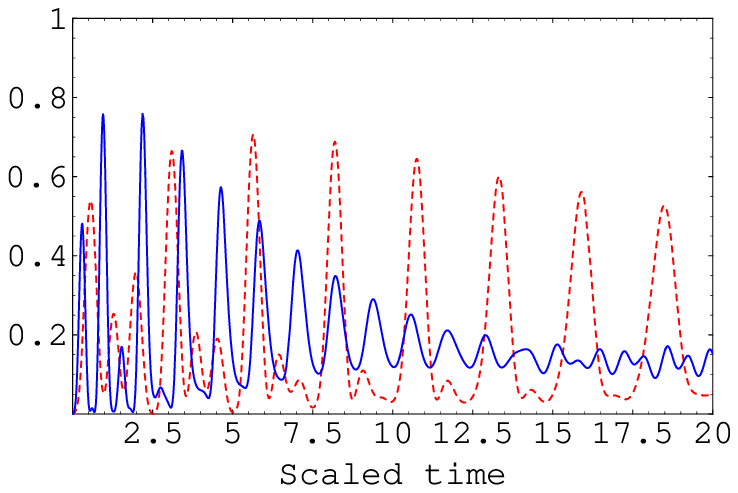} %
\includegraphics[width=18pc,height=12pc]{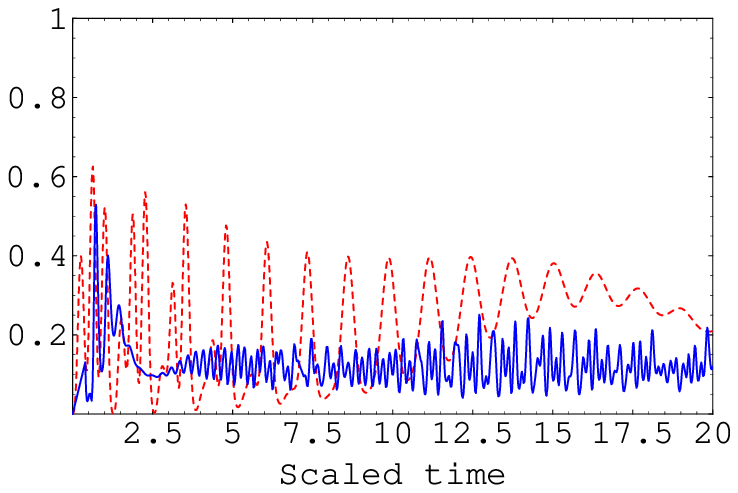}
\put(-430,80){$\mathcal{E}$} \put(-210,80){$\mathcal{E}$}
\end{center}
\caption{ The degree of entanglement $\mathcal{E}$ generated
between the two atoms prepared initially in the state
(\ref{InitAstate})
 The deformed parameter $q=0.5,0.9$
 and the mean photon numbers, $\bar n=10$ for the dot and solid curves respectively (a) $m=1$ and
(b)$m=2$. }
\end{figure}
Fig.(2a), shows the behavior of the degree of entanglement for
$m=1$. Since we start with a separable atomic system, the  degree
of entanglement, $\mathcal{E}=0$ at $\lambda t=0$. However, for
$\lambda t>0$,
 an entangled state generated between the two atoms. For $q=0.5$,
 one can  easily  see that the entanglement  of the two
atoms decreases as time increases( dot-curves). However,
$\mathcal{E}$ goes to zero for a very small  finite time and
suddenly increases.
 As one increases the deformity parameter ($q=0.9),$ the degree of
entanglement decreases gradually but does not vanish.   It is
clear that, the entanglement tends to zero several time for small
values of the deformity parameter. On the other hand, for larger
values of the deformity, $\mathcal{E}$ goes to zero only limited
time. The dynamics of the entanglement for  $m=2$ is displayed in
Fig.(2b), where it is smaller than that depicted in Fig.(1a).
However, the entanglement oscillates so fast as one increases the
value of the deformed parameter. It is clear that, the possibility
of obtaining a long lived entanglement is increases as one
increases the value of the deformed parameter, where we can see
that the entanglement is  zero only when $\lambda t=0.$

One can conclude that the number of photons plays  as  a control
parameter, where one can improve the degree of entanglement
between the two atoms, by increasing the number of photons inside
the cavity. Also, the long-lived entanglement between the two
atoms can be obtained by increasing the deformation.

 \section{Teleportation}

 In this section, the generated entangled state (\ref{FinalAtomic})
is employed to implement the original quantum teleportation
protocol \cite{ben}. This protocol is used to send unknown quantum
state between two users Alice and Bob, by using local operations
and classical communication. Now assume that, Alice is given
unknown state defined by,
\begin{equation}
\rho _{u}=\frac{1}{2}(1+\row{s_u}\cdot\sigma), \label{unknown}
\end{equation}
where the componnents of the Bloch vector $\row{s_u}$ are given by
\begin{equation}
s_{u_x}=\alpha \beta ^{\ast }+\beta \alpha ^{\ast
},s_{u_y}=i(\beta \alpha ^{\ast }-\alpha \beta ^{\ast
}),s_{u_z}=|\alpha |^{2}-|\beta |^{2},~ \mbox{where}
|\alpha|^2+|\beta|^2=1.
\end{equation}
To achieve the teleportation  Protocol, the partners follow the
following steps:
\begin{enumerate}
\item Alice performs the  CNOT gate on her qubit and the given
unknown qubit  followed by  Hadamard gate.

\item Alice measures her qubit and the unknown qubit randomly in one of the
basis $\left\vert ee\right\rangle ,\left\vert eg\right\rangle ,\left\vert
ee\right\rangle $ and \ $\left\vert gg\right\rangle $ and sends her results
to Bob by using classical channel.

\item As soon as Bob receives the classical data from Alice, he
applies a single qubit operation on his qubit depending on Alice's
results. So, if Alice measures in the basis  $\ket{ee}$, Bob will
obtain the state.
\begin{equation}
\rho _{Bob}=\frac{1}{2}(1+\row{s}_{b}\cdot\row\sigma),
\label{Rbob}
\end{equation}%
where, $s_{x_b},s_{y_b},s_{z_b}$ are the components of
$\row{s}_{b}$,
\begin{eqnarray}
s_{x_b} &=&|\alpha |^{2}(c_{n}^{(3)}c_{n-1}^{(\ast
4)}+c_{n}^{(4)}c_{n+2}^{(\ast 3)})+\alpha \beta ^{\ast
}(c_{n}^{(4)}c_{n-1}^{(\ast 1)}+c_{n}^{(3)}c_{n}^{(\ast 2)})-  \nonumber \\
&&\alpha ^{\ast }\beta (c_{n}^{(1)}c_{n-2}^{(\ast
2)}+c_{n}^{(3)}c_{n}^{(\ast 2)})+|\beta |^{2}(c_{n}^{(1)}c_{n-2}^{(\ast
2)}+c_{n}^{(2)}c_{n+2}^{(\ast 1)}),
 \nonumber \\
s_{y_b} &=&|\alpha |^{2}(c_{n}^{(4)}c_{n+2}^{(\ast
3)}-c_{n}^{(3)}c_{n-2}^{(\ast 4)})+\alpha \beta ^{\ast
}(c_{n}^{(4)}c_{n-1}^{(\ast 1)}-c_{n}^{(3)}c_{n}^{(\ast 2)})  \nonumber \\
&&-\alpha ^{\ast }\beta (c_{n}^{(3)}c_{n}^{(\ast
2)}-c_{n}^{(1)}c_{n-2}^{(\ast 2)})+|\beta |^{2}(c_{n}^{(2)}c_{n+2}^{(\ast
1)}-c_{n}^{(1)}c_{n-2}^{(\ast 2)}),
 \nonumber \\
s_{z_b} &=&|\alpha
|^{2}(|c_{n}^{(3)}|^{2}-|c_{n}^{(4)}|^{2})+\alpha \beta ^{\ast
}(c_{n}^{(3)}c_{n+2}^{(\ast 1)}-c_{n}^{(4)}c_{n+2}^{(\ast 2)})+
\nonumber \\
&&\alpha ^{\ast }\beta (c_{n}^{(2)}c_{n-2}^{(\ast
4)}-c_{n}^{(1)}c_{n-2}^{(\ast 3)})+|\beta
|^{2}(|c_{n}^{(1)}|^{2}-|c_{n}^{(2)}|^{2}).
\end{eqnarray}
\end{enumerate}
\begin{figure}
\begin{center}
\includegraphics[width=18pc,height=12pc]{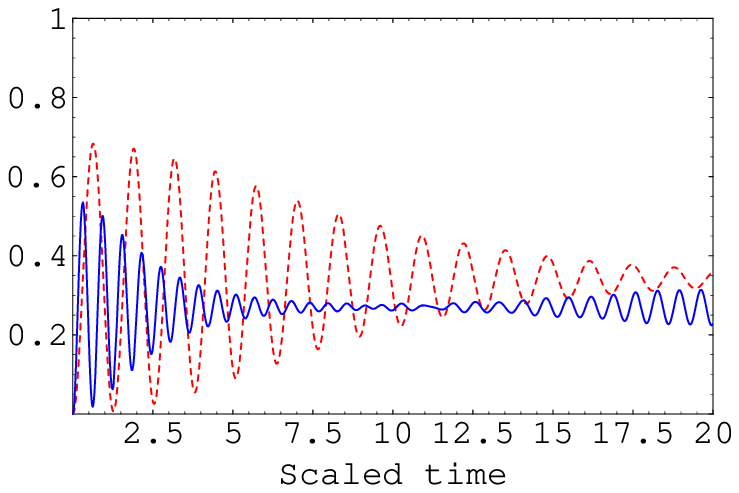} %
\includegraphics[width=18pc,height=12pc]{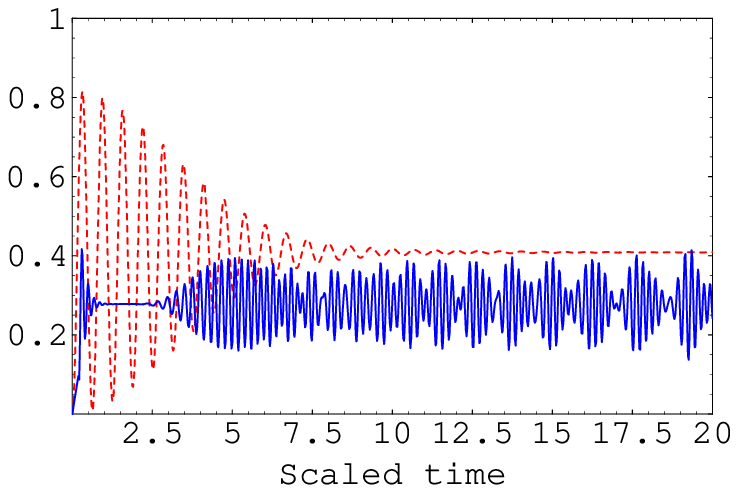}
\put(-430,80){$\mathcal{F}$} \put(-210,80){$\mathcal{F}$}
\end{center}
\caption{ The Fidelity of the teleported state (\ref{unknown})
with $s_{x_{u}}=1, s_{y_{u}}=s_{z_{u}}=0$, and the deformed
parameter $q=$0.5$, 0.9$ and $\bar n=10$ for the dot and solid
curves respectively (a) $m=1$ and (b)$m=2$. }
\end{figure}
The accuracy of the teleported   state  (\ref{unknown}) is
quantified  by evaluating the fidelity, $\mathcal{F}$
\begin{equation}
\mathcal{F}=\frac{1}{4}(1+\row{s_u}\cdot\row{s}_{Bob}).
\end{equation}
In Fig.(3), we investigate the dynamics of the fidelity
$\mathcal{F}$ for different values of the field and atomic
parameters.  We use the generated entangled state which is defined
by its Bloch vectors (\ref{FinalBloch}) and the cross dyadic
(\ref{Finaldyd}) with $a_1=1$ and $a_2=a_3=a_4=0$. Also, we assume
that Alice is given information  coded in the state
(\ref{unknown}) with $s_{x_u}=1$ and $s_{y_u}=s_{z_u}=0$.
Fig.(3a),  shows the behavior of $\mathcal{F}$, for $m=1$ and
different values of the deformity parameter. From this figure it
is clear that, for $q=0.5$, the amplitude of the oscillations of
the fidelity decreases as time increases. As one increases the
deformity parameter $(q=0.9)$, the fidelity  is smaller than
depicted for small values of $q$. Also, the phenomena of revivals
and collapse is clearly displayed for larger values of the
deformity parameter.

In Fig.(3b), we increase the number of photons inside the cavity,
($m=2$). In this case the  flocculations of the fidelity is very
fast for  small range of time. But as time increases these
revivals almost disappear and the fidelity is almost constant.
This behavior is changed dramatically as one increases the value
of the deformity parameter ($q=0.9)$, where the collapses appear
for small range of time. But as time goes on, the fidelity
oscillates so fast  but always the amplitudes of the fidelity
$\mathcal{F}$ is smaller than that shown in Fig.(3a). These
results show that Bennett's teleportation protocol \cite{ben} is
robust against the deformation

\section{Conclusion}
In  this contribution, a system of two atoms interacts with a
deformed cavity mode is introduced. The solution of the system is
introduced by means of the Bloch vectors  and the cross dyadic of
the travelling atoms inside the cavity. Due to the interaction,
there are  some entangled states have been generated. The amount
of entanglement between the entangled atoms is quantified by using
a measure
 depends on the Bloch vectors and the cross dyadic.
Although,  as one increases the deformity, the entanglement
decreases more, but it  is more robust, where it vanishes  only on
small interval of time. The deformation of entanglement appears
clearly as one increases the number of photons inside the cavity,
where the entanglement oscillates so fast in irregular behavior.
Also, we investigate the dynamics of the Bloch vectors, where we
show that as one increases the deformation parameter, the
amplitudes of the Bloch vectors decrease much faster. One can
obtain, entangled states of Werner types with high degree of
entanglement.

The effect of the deformity  on the accuracy  of the transported information between two users is investigated.
 The amplitude of the fidelity decreases  in the presence
of the deformity. This means that the minimum value of the
fidelity is improved as time goes on. For large values of the
deformed parameter, the fidelity decreases and the  oscillations
of the  amplitudes decrease more. If the number of photons inside
the cavity increases, the fidelity flocculates very fast in small
range of time and as time goes on the fidelity is almost constant.
The deformation of the fidelity  appears for larger values of the
deformity parameter, where the  revivals are very fast.

Finally, one can conclude that the deformation and the number of
photons
 inside the cavity play an important role on the dynamics of
entanglement, Bloch vectors and the fidelity of the teleported
state. If the devices are defective manufacturing, then by
increasing the number of photons within the cavity one  can reduce
the distortion resulting from such defects. The effect of
deformation  is different from other noise which leads to a sudden
death of entanglement and consequently a sudden death of
communication. So, we expect that these results are important in
quantum communication and consequently in building quantum
computers

{\bf Acknowledgment}

I would like to thank Prof. B-G. Englert his fruitful discussion
and the  important remarks which has improve the manuscript.

\end{document}